\documentclass[conference]{IEEEtran}
\usepackage{cite}
\usepackage{amsmath,amssymb,amsfonts}
\usepackage{algorithmic}
\usepackage{graphicx}
\usepackage{textcomp}
\usepackage{xcolor}
\usepackage{tikz}
\usepackage{pgfplots}
\usepackage[mode=buildnew]{standalone}
\pgfplotsset{compat=1.18}
\usepackage{gensymb}
\usepackage{booktabs}
\usepackage{comment}
\usepackage{lipsum}
\usetikzlibrary{fit,arrows,positioning,arrows.meta}
\usetikzlibrary{patterns}
\usetikzlibrary{shapes.arrows}
\usepackage{hyperref}
\usepackage{multirow}
\usepackage[T1]{fontenc}

\usepackage{titlesec}
   \titlespacing*{\section}      {0pt}{2ex plus 0ex minus 0ex}{1ex plus 0ex minus 0ex}%
 \titlespacing{\subsection}{0pt}{1pt plus 0pt minus 1pt}{0pt plus 0pt minus 1pt}
 \titlespacing{\subsubsection}{0pt}{1pt plus 0pt minus 5pt}{1pt}

\def\BibTeX{{\rm B\kern-.05em{\sc i\kern-.025em b}\kern-.08em
    T\kern-.1667em\lower.7ex\hbox{E}\kern-.125emX}}
\begin{document}
\title{Theoretical and Experimental Evaluation of AoA Estimation in Single-Anchor 5G Uplink Positioning}

\author{\IEEEauthorblockN{Thodoris Spanos$^{1,2}$, 
Fran Fabra$^3$, Jos\'e A. L\'opez-Salcedo$^3$, Gonzalo Seco-Granados$^3$, \\Nikos Kanistras$^2$, {Ivan Lapin$^4$}, Vassilis Paliouras$^1$
}\vspace{0.1cm}
\IEEEauthorblockA{$^1$Dept. of Electrical and Computer Engineering, University of Patras, Greece\\ $^2$Loctio, Greece\\
$^3$Dept. of Telecommunication and Systems Eng., Universitat Aut\`onoma de Barcelona (UAB), Spain\\
 $^4$Radio Navigation Systems and Techniques Section, European Space Agency, The Netherlands}
}

\maketitle

\begin{abstract}
As we move towards 6G, the demand for high-precision, cost-effective positioning solutions becomes increasingly critical. Single-anchor positioning offers a promising alternative to traditional multi-anchor approaches, particularly in complex propagation environments where infrastructure costs and deployment constraints present significant challenges. This paper provides a comprehensive evaluation of key algorithmic choices in the development of a single-anchor 5G uplink positioning testbed. Our developed testbed uses angle of arrival (AoA) estimation combined with range measurements from an ultra-wideband pair, to derive the position. The simulations conducted assess the impact of the selected algorithms on channel order and AoA estimation, while the influence of antenna calibration errors on AoA estimation is also examined. Finally, we compare simulations and results obtained from our developed platform.
\end{abstract}

\begin{IEEEkeywords}
SRS, 5G, AoA, DoA, MUSIC, ESPRIT, 2D ESPRIT, testbed, positioning, localization
\end{IEEEkeywords}

\section{Introduction}

The deployment of 5G networks has significantly enhanced connectivity by offering a wide range of capabilities such as faster speeds and lower latency. As the industry transitions to 6G, new services will demand ultra-high precision positioning~\cite{5gppp}. In addition, metrics such as reliability, availability, power consumption, and security will be fundamental design considerations~\cite{5gppp}. Positioning accuracies have reached impressive benchmarks---down to a few decimeters indoors and a few meters outdoors---surpassing earlier technologies such as Long Term Evolution (LTE) and even Global Navigation Satellite Systems (GNSS) in challenging indoor environments~\cite{pos_5g}. As the evolution towards 6G continues, further advancements in multi-antenna technologies\cite{5gantenna}, reconfigurable intelligent surfaces (RISs)~\cite{ris_wym}, and joint communication and sensing~\cite{5g_jcas} will continue to push the limits of localization accuracy.

Traditional localization techniques rely on multiple spatially distributed reference nodes to achieve high positioning accuracy~\cite{ita}. However, deploying multi-anchor setups is often impractical due to prohibitive infrastructure costs or physical constraints within the environment. Single-anchor positioning has emerged as an attractive solution, enabling location estimation with only one base station (BS). To accommodate positioning, the 5G standard defines specific positioning signals~\cite{3GPPsrs}. The Sounding Reference Signal (SRS) is utilized for positioning in the uplink (UL) although it was originally designed for channel estimation and beam management. Similarly, for downlink positioning, the Positioning Reference Signal (PRS) is used primarily. However, the limited commercial availability of PRS has led to the use of opportunistic signals for positioning, such as the Synchronization Signal Block~\cite{ssb1,ssb2}. 

Recent years have witnessed significant advances in testbed development aimed at validating 5G positioning technologies under realistic conditions. Wang \emph{et al.} demonstrated a 5G positioning testbed in an industrial indoor environment, using carrier-phase measurements to achieve centimeter-level positioning accuracy through differential phase difference of arrival techniques~\cite{wang}. Vordonis~\emph{et al.} apply a beam sweeping protocol utilizing an RIS prototype for the estimation of angle of arrival (AoA)~\cite{Vordonis}. Malm~\emph{et al.} developed an ultra-dense network (UDN) testbed designed to validate location-based handover mechanisms, effectively using UL-SRS for real-time AoA tracking~\cite{Malm}. Further work by Menta~\emph{et al.} addressed practical antenna array imperfections, such as mutual coupling and clock drift between Software Defined Radios front-ends, to enhance AoA-based positioning reliability in UDNs~\cite{Menta}. Li~\emph{et al.} propose a joint AoA and Time-of-Flight (ToF) method with a single base station, utilizing Channel State Information (CSI) \cite{Li}. Blanco \emph{et al.} developed an LTE localization testbed based on the direction and time of arrival~\cite{LTEtb}. 

In this paper, we present a comparison between simulation-based evaluations leading to key algorithmic decisions, and real-world measurements obtained using our fully developed single-anchor 5G-UL positioning testbed~\cite{spanos}. The remainder of the paper is organized as follows: Section \ref{testbed_sec} introduces the motivation for a single-anchor setup and describes our experimental testbed. Section \ref{System} details the system model and presents the localization algorithms along with their performance evaluation. In Section \ref{antenna_errors}, we quantify the impact of antenna-calibration errors. Section \ref{tb_res} reports the testbed results and compares them with simulation outcomes. Finally, Section \ref{conclusions} summarizes our conclusions.

\section{Testbed Description}
\label{testbed_sec}

The developed testbed~\cite{spanos} integrates two Universal Software Radio Peripherals (USRPs) for AoA estimation with an ultra-wideband module (UWB) for range measurements, enabling single-anchor positioning relative to the BS. A UWB ToF measurement constrains the UE location to a narrow circle, while the AoA estimate provides a bearing ray that resolves angular ambiguity. The very compact Ettus E312 serves as the transmitter, and an Ettus N310---with four Rx channels---serves as the receiver. The system transmits UL-SRS, which the N310 captures over 2.4 GHz, 3.5 GHz, and 5.8 GHz bands with 20 MHz and 50 MHz bandwidths. The receiver employs an---uncalibrated---uniform linear array (ULA) z omnidirectional antennas spaced at $\lambda$/2. Each is connected to an individual channel of the N310, with the fourth available channel reserved for calibration purposes~\cite{spanos}. 

Prior to hardware deployment, extensive simulations were performed to investigate the accuracy of the AoA estimation and algorithmic performance in controlled scenarios. These simulations provided valuable insights into system behavior, helping us to define key parameters and anticipate potential challenges. Upon completion of the testbed, a campaign of static and pedestrian measurements was conducted to validate its performance. Pedestrian test results based on raw measurement data, together with a detailed description of the testbed, are presented in~\cite{spanos}, while preliminary static results combined with a ray-tracing analysis are discussed in~\cite{spanos2}. Building upon this earlier work, the current study provides a thorough evaluation of the employed signal processing techniques and investigates how antenna imperfections influence overall system performance. Additionally, comparisons between simulation outcomes and recent static test results, extending beyond those reported in~\cite{spanos}, are also presented.
\section{System Model and Algorithmic Decisions}
\label{System}
Simulations model a static UE and a BS equipped with a three-element ULA positioned 20 m apart in a static position, at a configurable angle. A 5G slot containing 14 OFDM symbols---with the SRS occupying the first four symbols in a comb type $K_\text{TC} = 2$ pattern---is continuously transmitted in the time domain. The BS detects the slot start, estimates the number of received signals, and computes the AoA. 

The SRS sequence is mapped onto the resource element grid as in~\cite{3GPPsrs}. The results of this work employ a waveform configuration with numerology $\mu = 1$ and 50 MHz bandwidth. After the subcarrier allocation, the SRS sequence is transformed in the time domain and transmitted through an Urban Micro channel model with added AWGN. To emulate signals arriving from multiple directions, additional signal components were synthesized by modifying the line-of-sight (LOS) component with different attenuation levels and propagation delays.

Assuming an $M$-element ULA, $M$ copies of the transmitted signal are received, one per antenna element, having the form
\begin{equation}
    \mathbf{y}(t) = \sum_{p=1}^{P} \mathbf{A}(\theta_p)\mathbf{H_p}(t) \mathbf{s}(t-\tau_p) + \mathbf{n}(t),
    \label{signal_reception}
\end{equation}
where $P$ is the total number of paths, $\mathbf{H_p}(t)$ is the path loss of each multipath component (MPC), $\mathbf{s}(t)$ is the transmitted SRS sequence in the time domain delayed by $\tau_p$, $\mathbf{n}(t)$ is the AWGN, and $\mathbf{A}(\theta_p$) is the steering vector of the $p$-th path, 
\begin{equation} 
\mathbf{A}(\theta_p)=\left[\exp\left(-j\frac{2\pi f_c d (m-1)\sin\theta_p}{c}\right)\right]_{m=1}^M,
\label{steering}
\end{equation}
where $f_c$ is the carrier frequency, $d$ the element spacing, $c$ the speed of light, and $\theta_p$ the azimuth angle of path $p$. 

\subsection{Timing Synchronization}
The receiver computes the offset of the received waveform in samples using cross-correlation, as it is stored locally and unaffected by noise. The offset, relative to the original waveform, is determined as the index $k^*$ corresponding to the largest peak of the cross-correlation output $\textbf{h}(k)$, i.e., $k^* = \operatorname*{argmax}_{k} \mathbf{h}(k),$ where $\mathbf{h}(k) = \sum_{i=1}^{L_{\text{seq}} -1} \mathbf{s}^*(i) \mathbf{y}(k+i),$
$L_{\text{seq}}$ denotes the length of the transmitted waveform in samples and $\textbf{s}^*$ denotes the conjugate of $\textbf{s}$.

\subsection{Estimating the Received Number of Signals}
The super-resolution algorithms, i.e., Multiple Signal Classification (MUSIC), Estimation of the Signal Parameters via Rotational Invariance Techniques (ESPRIT), require a knowledge of the received number of signals. Several previous studies have addressed this topic \cite{Ivan}, \cite{Cheng}. The algorithms perform poorly when the number of signals received is not correctly estimated.
In the following, the Akaike Information Criterion (AIC) \cite{Akaike} and the Minimum Description Length (MDL)~\cite{MDL} techniques have been evaluated for channel order estimation.

The AIC is often used 
to estimate the channel order, $k$, in a mixture of signals (i.e., LOS path, MPC, other sources). The AIC value deduced from $N$ samples is calculated for each possible $k$, subsequently selecting $k^*_{\text{AIC}}$, i.e., the candidate number of signals with the lowest AIC value, as the most appropriate estimator of the actual number, as described by
\begin{equation}
    k^*_{\text{AIC}} = \operatornamewithlimits{argmin}\limits_n\big[-2N(M-k) \log L(k) + 2k(2M-k)\big],
    \label{Akaike}
\end{equation}
where $L(k)$ is a log-likelihood function used to estimate the maximum likelihood of signal number $k$, defined in \cite{Cheng} as
$L(k) = \frac{{(z_{k+1} z_{k+2} \ldots z_{M})}^{\frac{1}{M-k}}}{\frac{1}{M-1}({z_{k+1} + z_{k+2} + \ldots + z_{M})}},$ and $z_1\geq z_2 \geq \ldots \geq z_M$ are the eigenvalues of the covariance matrix $\mathbf{R} \in \mathbb{C}^{M \times M}$.

The MDL principle is another widely used model selection criterion in statistical modeling and hypothesis testing. MDL calculates the coding length required to describe each candidate, taking into account both the parameters and the data. The candidate with the shortest coding length is selected as the most appropriate one, as described by
\begin{align}
     k^*_{\text{MDL}} = \operatornamewithlimits{argmin}_k\Big[{-}2N(M{-}k)&\log L(k) &
     \nonumber\\
     & +\frac{k}{2}(2M{-}k)\!\log N\Big ].
     \label{MDL}
\end{align}
\begin{figure}[h]
\begin{tikzpicture}[scale=1, transform shape]
\begin{axis}[
    legend style={
        at={(1.,1)},
        anchor=north west,
        nodes={scale=0.65, transform shape},
        font=\normalsize
    },
    height=5.2cm,
    width=7cm,
    xlabel=SNR (dB), ylabel=Hits ($\%$), 
        ytick={0,20,40,60,80,100},
        yticklabels={0,20,40,60,80,100},xmin=-5, xmax=10, ymin=0, ymax=100, 
    grid=both,
    minor x tick num=3
]
    \addplot[red,mark=square, mark size=1.5pt] table[x index = 0, y index = 1 ]{COE/3_elements.tex};
    \addplot[dashed,red,mark=none, mark size=1.5pt] table[x index = 0, y index = 2 ]{COE/3_elements.tex};
    \addplot[dashdotted,red,mark=none] table[x index = 0, y index = 3 ]{COE/3_elements.tex};
    \addplot[red,mark=x, mark size=1.5pt] table[x index = 0, y index = 4 ]{COE/3_elements.tex};
    \addplot[red,mark=o, mark size=1.5pt] table[x index = 0, y index = 5 ]{COE/3_elements.tex};
    \addplot[black,mark=square, mark size=1.5pt] table[x index = 0, y index = 6 ]{COE/3_elements.tex};
    \addplot[dashed,black,mark=none] table[x index = 0, y index = 7 ]{COE/3_elements.tex};
    \addplot[dashdotted,black,mark=none] table[x index = 0, y index = 8 ]{COE/3_elements.tex};
    \addplot[black,mark=x, mark size=1.5pt] table[x index = 0, y index = 9 ]{COE/3_elements.tex};
    \addplot[black,mark=o, mark size=1.5pt] table[x index = 0, y index = 10 ]{COE/3_elements.tex};

    \addlegendentry{Akaike 10\textdegree}
    \addlegendentry{Akaike 8\textdegree}
    \addlegendentry{Akaike 6\textdegree}
    \addlegendentry{Akaike 4\textdegree}
    \addlegendentry{Akaike 2\textdegree}
    \addlegendentry{MDL 10\textdegree}
    \addlegendentry{MDL 8\textdegree}
    \addlegendentry{MDL 6\textdegree}
    \addlegendentry{MDL 4\textdegree}
    \addlegendentry{MDL 2\textdegree}
\end{axis}
\end{tikzpicture}
\caption{AIC and MDL comparison: LOS-MPC angle separation 2\degree{}--10\degree.}
\label{Ak v MDL}
\end{figure}
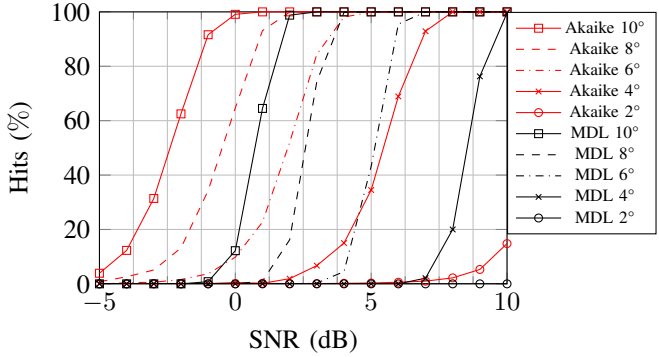
Both criteria were evaluated for channel order estimation, in terms of accurately estimating the number of closely-spaced MPCs. A ``hit'' occurs when the estimated number of signals $k^*$ coincides with the actual scenario, otherwise it is a ``miss''. Fig.~\ref{Ak v MDL} compares the AIC and MDL estimates over 1000 Monte-Carlo simulations, corresponding to a scenario of a LOS path set at 0\degree \space plus one MPC of $\frac{7}{10}$ the LOS power, separated by 2\degree{}--10\degree{} from the LOS path. AIC outperforms MDL at all five angle separations for the three-element ULA setup, with an average of 2 dB. Based on these results, AIC was used for channel-order estimation in our testbed.

\subsection{AoA Estimation}
We considered three well-known super-resolution AoA estimation algorithms for our testbed: MUSIC \cite{MUSIC}, ESPRIT \cite{ESPRIT} and 2D ESPRIT \cite{2DESPRIT}. 
MUSIC is a well-established super-resolution direction-finding algorithm based on the eigenvalue decomposition of $\mathbf{R}$. It detects up to $WM-1$ signals by decomposing $\mathbf{R}$ into the signal subspace $\mathbf{U}_s = [\mathbf{v}_1, \dots, \mathbf{v}_k]$ and the noise subspace $\mathbf{U}_n = [\mathbf{u}_{k+1}, \dots, \mathbf{u}_M]$. Since $\mathbf{R}$ is Hermitian, these subspaces are orthogonal, and MUSIC exploits this via the spectrum
\begin{equation}
    \text{MUSIC}_{\text{Spectrum}} = \frac{1}{\mathbf{A}^H \mathbf{U}_n \mathbf{U}_n^H \mathbf{A}}.
\end{equation}
ESPRIT offers an efficient alternative that takes advantage of the rotational relationship between subarrays. It divides the antenna array into two overlapping subarrays $\mathbf{A_1}$ and $\mathbf{A_2}$, with $\mathbf{A_2} =  \mathbf{A_1 \Xi}$, where $\mathbf{\Xi}$ is a diagonal matrix with entries  
\begin{equation}   
\xi_i = \exp\left(\frac{-2j \pi d \sin{\theta_i}}{\lambda}\right),
\end{equation}
where $\theta_i$ is the AoA, and $\lambda$ the wavelength. ESPRIT exploits this rotational structure in the signal subspace $\mathbf{U_s}$, forming $\mathbf{S} = \mathbf{U_s}(:,1:k), \label{e:SU}$
where $\mathbf{S}$ contains the first $k$ columns of $\mathbf{U_s}$. A matrix $\mathbf{P}$ relates subspaces $\mathbf{S_2} = \mathbf{S_1} \mathbf{P}$, estimated via Least Squares as $\mathbf{P} = \mathbf{S}_1^* \mathbf{S}_2/{\mathbf{S}_1^* \mathbf{S}_1}.$ The angle $\theta$ is estimated as $\theta  = \arcsin(\rho)$, where $\rho = \phi_i/{2\pi d},$ and $\phi_i$ is the $i$-th phase of the $I$ eigenvalues of $\mathbf{P}$.

The 2D ESPRIT algorithm extends ESPRIT by jointly estimating both angles and delays. 2D ESPRIT forms a Hankel matrix by stacking copies of the CSI matrix \(\mathbf{H}\), exposing its shift-invariant structure. Similar to the 1D case, to compute the delays $\tau_i$, a diagonal matrix $\mathbf{\Psi}$ with entries 
\begin{equation}
\psi_i = \exp\left(\frac{-2j \pi \tau_i}{l}\right),
\end{equation}
is defined, where $l$ is the channel length (in symbol periods). The data model is $\mathbf{H} = \mathbf{A B F},$ where $\mathbf{A}$ is the Khatri-Rao product of the steering and delay matrices, $\mathbf{B}$ contains path attenuations, and $\mathbf{F}$ is a Vandermonde DFT matrix. Selection matrices extract angles $\xi_i$ and delays $\psi_i$, and the Vandermonde structure of $\mathbf{F}$ allows joint diagonalization to correctly pair each angle with its corresponding delay. 

\begin{figure}
\begin{tikzpicture}
\begin{axis}[legend style={
    at={(1,0.52)},
    anchor=south east,
    nodes={scale=0.65, transform shape}
},  height = 6cm, width = 8.7cm,
        xlabel=SNR (dB), ylabel=RMSE (\degree), xmin=-5, xmax=10, ymin=0, 
        ytick={0,0.1,...,0.8},
        yticklabels={0\degree{},,0.2\degree{},,0.4\degree{},, 0.6\degree{}, , 0.8\degree{}},
        ymax=0.8,
        grid=both,
        minor x tick num=3]

    \addplot[teal,mark=square, mark size=1.5pt,mark repeat = 2, mark phase = 2] table[x index = 0, y index = 1 ]{Data/0los.tex};
    \addplot[blue,mark=square, mark size=1.5pt,mark repeat = 2, mark phase = 1] table[x index = 0, y index = 2 ]{Data/0los.tex};
    \addplot[red,mark=square, mark size=1.5pt,mark repeat = 2, mark phase = 2] table[x index = 0, y index = 3 ]{Data/0los.tex}; 
    \addplot[teal,mark=o, mark size=1.5pt,mark repeat = 2, mark phase = 2] table[x index = 0, y index = 3 ]{Data/mpc_15.tex}; 
    \addplot[blue,mark=o, mark size=1.5pt,mark repeat = 2, mark phase = 1] table[x index = 0, y index = 2 ]{Data/mpc_15.tex}; 
    \addplot[red,mark=o, mark size=1.5pt, mark repeat = 2, mark phase = 1] table[x index = 0, y index = 1]{Data/mpc_15.tex}; 
    \addplot[black,mark=none] table[x index = 0, y index = 4 ]{Data/0los.tex}; 

    \addlegendentry{MUSIC - No MPC}
    \addlegendentry{ESPRIT - No MPC}
    \addlegendentry{2D ESPRIT - No MPC}
    \addlegendentry{MUSIC - 15\degree{} MPC}
    \addlegendentry{ESPRIT - 15\degree{} MPC}
    \addlegendentry{2D ESPRIT - 15\degree{} MPC}
    \addlegendentry{Cramer-Rao Bound}

\end{axis}
\end{tikzpicture}
\caption{Comparison between MUSIC, ESPRIT and 2D ESPRIT.}
\label{music vs esprit}
\end{figure}
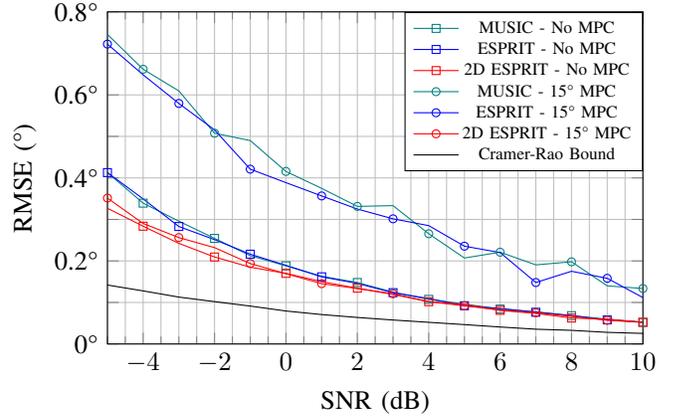

The performance of the three AoA algorithms is compared in Fig.~\ref{music vs esprit} for two scenarios: one with a single LOS path at~$0\degree{}$, and another with an additional MPC separated by~$15\degree{}$ from the LOS. Results were averaged over 1000 Monte Carlo simulations, assuming a perfect channel order estimation. Additionally, the Cramér-Rao bound (CRB) for the LOS-only scenario was derived as a theoretical lower bound.

In both cases, MUSIC and ESPRIT performed similarly, approaching the CRB as the signal-to-noise ratio (SNR) increased, aligning with the theoretical expectations. 2D ESPRIT achieved higher accuracy than conventional 1D methods, primarily due to its utilization of the CSI, albeit at a higher computational cost. Considering the need for processing speed, ESPRIT was chosen as the primary algorithm for AoA estimation because of its closed-form solution and lower complexity, although the remaining evaluated algorithms remain supported.

\section{Antenna Calibration Errors}
\label{antenna_errors}
The results presented in Section \ref{System} assume ideal conditions, ignoring hardware imperfections such as uneven antenna gains, phase mismatches, mutual coupling, orientation, and geometrical misalignments---all of which can degrade the accuracy of AoA estimation in real-world conditions. Given that our testbed employs an uncalibrated antenna array, it is essential to account for these factors. In this section, we investigate the effects of antenna calibration errors on AoA estimation, providing theoretical insights supported by simulation results. Note that while antenna orientation and element spacing errors are also important, they are beyond the scope of this analysis.

In a more generalized expression than \eqref{steering}, the $m$-th antenna element of the steering vector $\mathbf{A}(\theta)$, with $1\leq m \leq M$, for an impinging angle $\theta$ is expressed as
\begin{equation}
[A(\theta)]_m = g_m e^{j\theta_m},
\end{equation}
where $g_m$ and $\theta_m$  represent the element's nominal gain and phase delay. Due to non-ideal factors, i.e., hardware imperfections, a new steering vector $\mathbf{A'}(\theta) = \mathbf{P} \cdot \mathbf{A}(\theta)$, is defined
\begin{align}
[A'(\theta)]_m &= (g_m + \Delta g_m)e^{j(\theta_m + \Delta \theta_m)}\notag \\ 
&= [A(\theta)]_m \left(1 + \frac{\Delta g_m}{g_m}\right) e^{j\Delta \theta_m} = \mathbf{P} \cdot \mathbf{A}(\theta).
\end{align}

Mutual coupling effects are introduced through a coupling matrix $\mathbf{C}$, applied to the perturbed steering vector, yielding
\begin{equation}
\mathbf{A''}(\theta) = \mathbf{C} \cdot \mathbf{A'}(\theta) = \mathbf{C} \cdot \mathbf{P} \cdot \mathbf{A}(\theta),
\end{equation}
where the coupling matrix $\mathbf{C}$ is formed as follows
\begin{equation}
\mathbf{C} = \begin{bmatrix}
c_{1,1} & c_{1,2} & \cdots & c_{1,M} \\
c_{2,1} & \ddots  & \cdots & c_{2,M} \\
\vdots  &         & \ddots & \vdots  \\
c_{M,1} & c_{M,2} & \cdots & c_{M,M}
\end{bmatrix}.
\end{equation}
The complex weight $c_{q,m}$ quantifies the coupling effect of the of the $m$-th antenna onto the $q$-th array element assumed independent of the signal's direction of arrival (DOA). Mutual coupling affects off-diagonal terms ($m \ne q$) in $\mathbf{C}$. Diagonal terms account for intrinsic perturbations such as group delay. 

Some studies \cite{ana} simplify antenna calibration by combining all hardware imperfections, including mutual coupling and phase center variations, into a single coupling matrix $\mathbf{C}$, removing matrix $\mathbf{P}$. In contrast, other approaches \cite{dan} separate these imperfections into constant errors represented by $\mathbf{C}$ and DOA-dependent variations included in $\mathbf{P}$. 
Using the latter approach, \eqref{signal_reception} is modified as
\begin{equation}
    \mathbf{\tilde{y}}(t) = \sum_{p=1}^{P} \mathbf{C} \cdot \mathbf{P} \cdot \mathbf{A}(\theta_p)\mathbf{H_p}(t) \mathbf{s}(t-\tau_p) + \mathbf{n}(t),
\end{equation}
where $\mathbf{\tilde{y}}(t)$ is the received signal considering imperfections at the received signal.
Errors in uncalibrated antenna arrays fall into two categories. The first includes uncertainties in antenna positioning, orientation, and inherent antenna imperfections, captured by matrix $\mathbf{P}$. The second covers constant mismatches, such as variations in wiring lengths and delays, as well as mutual coupling effects, all incorporated within matrix $\mathbf{C}$.

The first group of errors are usually modeled as random variables following uniform distributions. In the latter case, the antenna gain and phase perturbations are modeled as
\begin{align}
\left(1 + \frac{\Delta g_m}{g_m}\right) & \sim u\left(10^{-P_{\text{gain}}/20} - 1, 10^{P_{\text{gain}}/20} - 1\right),
\label{amp_err} \\
\Delta\theta_m & \sim u\left(-P_{\text{phase}}\frac{\pi}{180}, P_{\text{phase}}\frac{\pi}{180}\right),
\label{phase_err}
\end{align}
where $u(x,y)$ denotes the uniform distribution probability density function in the range of $(x, y)$, while $P_{\text{gain}}$ and $P_{\text{phase}}$ denote gain and phase perturbations in dB and degrees, respectively. $P_{\text{gain}}$ is typically assumed to induce gain variations of up to 5\% relative to unity \cite{ebe,Dai} and is generally constrained to no more than 1 dB. For phase, Backen~\emph{et al.} demonstrate a GNSS antenna array with phase center variations of 1 cm---approximately 19\degree{}---which reflects the performance of low-cost, suboptimal antennas~\cite{Backen2008}. Similarly, He~\emph{et al.} employ a model where the antenna gain and phase are perturbed by independent zero-mean Gaussian random variables with standard deviations of 5\% and 5\degree{}, respectively~\cite{He2021}. Following the $3\sigma$ rule, these models allow for gain variations up to 15\% (roughly 0.7 dB degradation) and phase deviations up to 15\degree{}.

Mutual coupling errors can be modeled through electromagnetic simulations using the numerical electromagnetic code~\cite{bur}, or by assigning zero-mean Gaussian distribution values, as in \cite{ana,fri}. Alternatively, \cite{ebe} suggests approximating coupling coefficients using the Friis equation, based on the antennas' radiation patterns and inter-element distances, as
\begin{align}
|c_{q,m}| &\approx \sqrt{\frac{G_q(\theta_m)G_m(\theta_q)\lambda^2}{(4\pi d_{q,m})^2}} ,
\label{friis} \\
\arg(c_{q,m}) &\approx -\frac{2\pi d_{q,m}}{\lambda},
\label{friisphase}
\end{align}
for amplitude and phase respectively, where $G_q(\theta_m)$ is the gain of the $m$-th antenna in the direction to the $q$-th antenna, $d_{q,m}$ is the distance separating the $q$-th and $m$-th antennas. 
Then, the coefficients $c_{q,m}$ of the matrix $\mathbf{C}$ would become
\begin{equation}
c_{q,m} \approx |c_{q,m}| e^{j\arg(c_{q,m})},
\label{friis2}
\end{equation}
In this work, amplitude errors are modeled using \eqref{amp_err} with $P_{\text{gain}}$ values of 0 dB--0.7 dB, and phase errors are represented by \eqref{phase_err} with $P_{\text{phase}}$ spanning from $-$15\degree{} to 15\degree{}. The amplitude gains of the coupling coefficients are derived via~\eqref{friis}, while phase variations are modeled using a zero-mean Gaussian with variance $\pi/20$, as \eqref{friis2} yields unrealistic results. Note that since our testbed uses omnidirectional antennas, the factors $G_q(\theta_m)$ and $G_m(\theta_q)$ in~\eqref{friis} are set to 1. 
\begin{figure}
\begin{tikzpicture}
\begin{axis}[
    legend style={
        at={(1,1)},
        anchor=north west,
        nodes={scale=0.65, transform shape},
        font=\normalsize
    },
    height=5cm,
    width=6cm,
    xlabel=SNR (dB), ylabel=RMSE (\degree), xmin=-5, xmax=10, ymin=0, 
    ytick={0,0.5,...,2},
    yticklabels={0\degree{},0.5\degree{},1\degree{},1.5\degree{},2\degree{}},
    ymax=2.2,
    grid=both,
    minor x tick num=3
]
    \addplot[teal,mark=square, mark size=1.5pt, mark repeat = 3, mark phase = 3] table[x index = 0, y index = 1 ]{Data/antenna_err.tex}; 
    \addplot[blue,mark=square, mark size=1.5pt, mark repeat = 3, mark phase = 2] table[x index = 0, y index = 2 ]{Data/antenna_err.tex}; 
    \addplot[red,mark=square, mark size=1.5pt, mark repeat = 3, mark phase = 1] table[x index = 0, y index = 3 ]{Data/antenna_err.tex};  
    \addplot[teal,mark=o, mark size=1.5pt, mark repeat = 3, mark phase = 3] table[x index = 0, y index = 4 ]{Data/antenna_err.tex}; 
    \addplot[blue,mark=o, mark size=1.5pt, mark repeat = 3, mark phase = 2] table[x index = 0, y index = 5 ]{Data/antenna_err.tex}; 
    \addplot[red,mark=o, mark size=1.5pt, mark repeat = 3, mark phase = 1] table[x index = 0, y index = 6 ]{Data/antenna_err.tex}; 
    \addplot[teal,mark=x, mark size=1.5pt, mark repeat = 3, mark phase = 3] table[x index = 0, y index = 7 ]{Data/antenna_err.tex}; 
    \addplot[blue,mark=x, mark size=1.5pt, mark repeat = 3, mark phase = 2] table[x index = 0, y index = 8 ]{Data/antenna_err.tex}; 
    \addplot[red,mark=x, mark size=1.5pt, mark repeat = 3, mark phase = 1] table[x index = 0, y index = 9 ]{Data/antenna_err.tex}; 

    \addlegendentry{MUSIC - \textbf{P}}
    \addlegendentry{ESPRIT - \textbf{P}}
    \addlegendentry{2D ESPRIT - \textbf{P}}
    \addlegendentry{MUSIC - \textbf{C}}
    \addlegendentry{ESPRIT - \textbf{C}}
    \addlegendentry{2D ESPRIT - \textbf{C}}
    \addlegendentry{MUSIC - \textbf{P} \& \textbf{C}}
    \addlegendentry{ESPRIT - \textbf{P} \& \textbf{C}}
    \addlegendentry{2D ESPRIT - \textbf{P} \& \textbf{C}}
\end{axis}
\end{tikzpicture}
\caption{Antenna calibration errors impact (matrices \textbf{P}, \textbf{C}) on AoA estimation.}
\label{cal_err}
\end{figure}
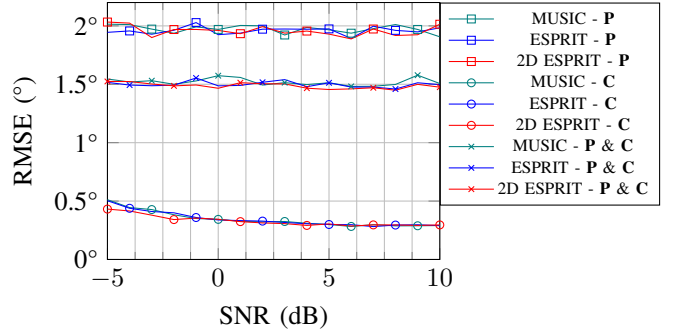
Fig.~\ref{cal_err} shows the impact of various calibration errors on AoA estimation for the case of a $0\degree{}$ LOS component over 1000 Monte Carlo simulations. We evaluate the performance of MUSIC, ESPRIT and 2D ESPRIT. First, we examine the impact of each matrix ($\mathbf{P, C}$) on the nominal steering vector independently, and then we assess their combined effect. The AoA algorithms use the nominal steering vector during the estimation process.

As seen in Fig.~\ref{cal_err}, introducing gain and phase mismatches alone (matrix $\mathbf{P}$) results in a consistent error of about $2\degree{}$, independent of SNR, and similar across all three algorithms. This suggests that $\mathbf{P}$ introduces systematic biases in the steering vector, which persist even at high SNR due to the fundamental model mismatch. Furthermore, when only mutual coupling errors (matrix $\mathbf{C}$) are present, the RMSE starts higher at low SNR but decreases as the SNR improves. This indicates that coupling primarily contributes to estimation variance rather than a fixed bias, allowing performance to improve as noise diminishes. When both $\mathbf{P}$ and $\mathbf{C}$ are introduced, the RMSE stabilizes at an intermediate value ($1.5\degree{}$). Interestingly, the combined scenario yields better performance than using matrix $\mathbf{P}$ alone, as mutual coupling errors introduce relatively structured perturbations that partially offset the otherwise random and systematic distortions caused by gain and phase mismatches. Although mutual coupling does not completely eliminate systematic biases from $\mathbf{P}$, it provides a stabilizing effect by introducing predictable coupling patterns, thus slightly improving overall estimation accuracy compared to the scenario with only gain and phase errors.

\section{Testbed Results}
\label{tb_res}

\begin{figure}
    \centering

    \begin{minipage}{\linewidth}
        \centering
        \includegraphics[width=\linewidth]{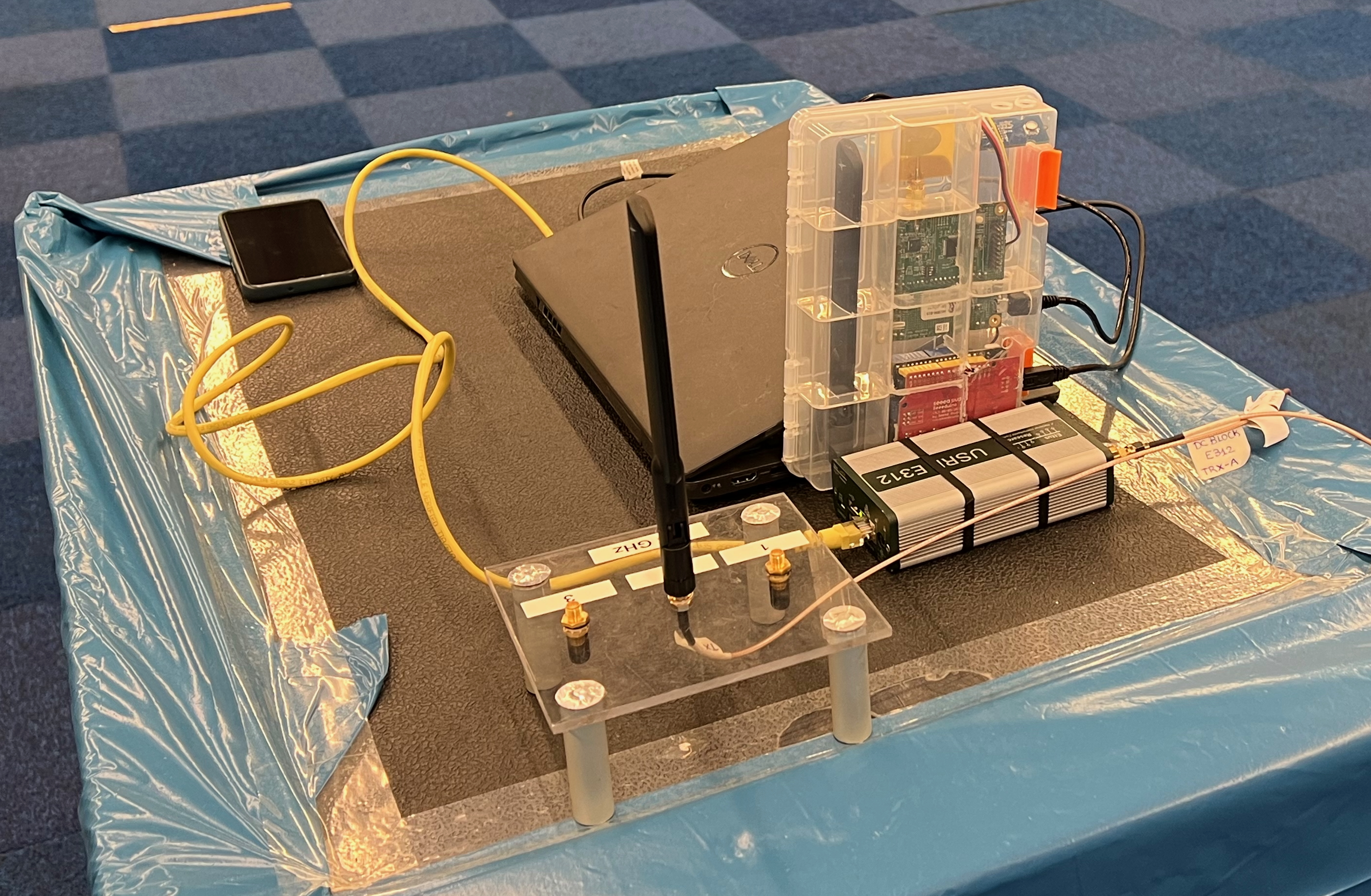}
    \end{minipage}
    
    \vspace{0.2cm}
    
    \begin{minipage}{\linewidth}
        \centering
        \includegraphics[width=\linewidth]{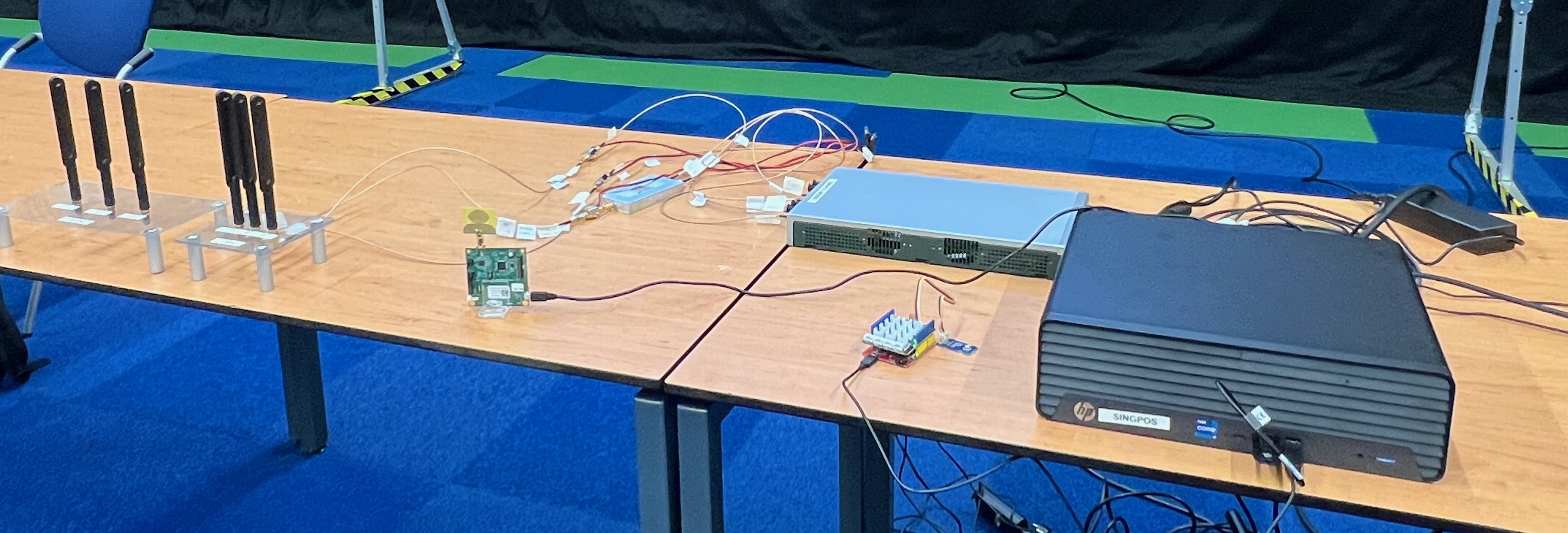}
    \end{minipage}
    \vspace{0.2cm}
    \caption{UE (above) and BS (below) of the developed testbed.}
    \label{testbed}
\end{figure}

Figure~\ref{testbed} illustrates the units employed as the UE and BS. More comprehensive details regarding the testbed design and key decisions taken concerning the hardware components, as well as preliminary results for a static UE and the full results obtained with a moving UE are provided in \cite{spanos, spanos2}.

In the experimental setup, a stored SRS slot is replayed from the Ettus E312 in an open field was captured using a three-element ULA connected to an Ettus N310. To account for long distances, low-noise amplifiers enhanced the received signal. The reception followed the procedure described in Section~\ref{System}. The UE remained static for 80 seconds, transmitting in the ISM 2.4 GHz band at a 0\degree{} angle relative to the BS and at the same height. To reduce IQ sample storage, a snapshot technique captured 6 ms of signal every second, with AoA estimates averaged over the number of slots contained in the received portion of the signal to yield one estimate per second.

\begin{figure}[!tb]
\centering
\begin{tikzpicture}
\begin{axis}[legend style={
    at={(-0.001,0.335)},
    anchor=south west,
    nodes={scale=0.8, transform shape}
},
        height = 6.5cm,
        width = 8.7cm,
        xlabel=Time (s), ylabel=Estimated AoA (\degree), xmin=1, xmax=80, ymin=-10, 
        ytick={-10,0,...,50},
        yticklabels={-10\degree{},0\degree{},10\degree{},20\degree{},30\degree{},40\degree{},50\degree{}},
        ymax=50,
        grid=both,
        minor x tick num=3]
    \addplot[teal,mark=square, mark size=1pt,mark repeat=6, mark phase=6, very thin] table[x index = 0, y index = 1 ]{Data/football_field.tex};
    \addplot[blue,mark=square, mark size=1pt,mark repeat=6, mark phase=4, very thin] table[x index = 0, y index = 2 ]{Data/football_field.tex};
    \addplot[red,mark=square, mark size=1pt,mark repeat=6, mark phase=2, very thin] table[x index = 0, y index = 3 ]{Data/football_field.tex}; 
    \addplot[teal,mark=o, mark size=1pt,mark repeat=6, mark phase=6, very thin] table[x index = 0, y index = 4 ]{Data/football_field.tex};
    \addplot[blue,mark=o, mark size=1pt,mark repeat=6, mark phase=4, very thin] table[x index = 0, y index = 5 ]{Data/football_field.tex};
    \addplot[red,mark=o, mark size=1pt,mark repeat=6, mark phase=2, very thin] table[x index = 0, y index = 6 ]{Data/football_field.tex}; 
    \addplot[teal,mark=x, mark size=1.5pt,mark repeat=6, mark phase=6, very thin] table[x index = 0, y index = 7 ]{Data/football_field.tex};
    \addplot[blue,mark=x, mark size=1.5pt,mark repeat=6, mark phase=4, very thin] table[x index = 0, y index = 8 ]{Data/football_field.tex};
    \addplot[red,mark=x, mark size=1.5pt,mark repeat=6, mark phase=2, very thin] table[x index = 0, y index = 9 ]{Data/football_field.tex}; 

    \addlegendentry{MUSIC - 15 m}
    \addlegendentry{ESPRIT - 15 m}
    \addlegendentry{2D ESPRIT - 15 m}
    \addlegendentry{MUSIC - 50 m}
    \addlegendentry{ESPRIT - 50 m}
    \addlegendentry{2D ESPRIT - 50 m}
    \addlegendentry{MUSIC - 80 m}
    \addlegendentry{ESPRIT - 80 m}
    \addlegendentry{2D ESPRIT - 80 m}

\end{axis}
\end{tikzpicture}
\caption{AoA estimation in time for 15 m, 50 m and 80 m.}
\label{AOA_TB}
\end{figure}
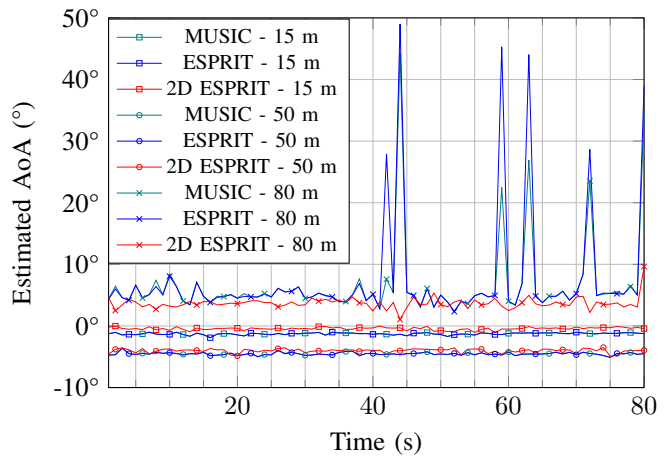
\begin{figure}[!tb]
\centering
\begin{tikzpicture}
\begin{axis}[legend style={
    at={(1,0)},
    anchor=south east,
    nodes={scale=0.8, transform shape}
},        
        height = 6.5cm,
        width = 8.9cm,
xlabel=AoA Error (\degree), ylabel={\shortstack{Cumulative Density Function}}, xmin=0, xmax=10, ymin=0, ymax=1,ytick={0,0.1,0.2,0.3,0.4,0.5,0.6,0.7,0.8,0.9,1},
    yticklabels={0,0.1,0.2,0.3,0.4,0.5,0.6,0.7,0.8,0.9,1},    grid=both,minor x tick num=3,
    unbounded coords=jump]
    \addplot[teal,mark=square, mark size=1.5pt,mark repeat=6, mark phase=6, very thin] table[x index = 1, y index = 0 ]{Data/cdfs/15m_music.tex};
    \addplot[blue,mark=square, mark size=1.5pt,mark repeat=6, mark phase=4, very thin] table[x index = 1, y index = 0 ]{Data/cdfs/15m_esprit.tex};
    \addplot[red,mark=square, mark size=1.5pt,mark repeat=6, mark phase=2, very thin] table[x index = 1, y index = 0 ]{Data/cdfs/15m_2desprit.tex}; 
    \addplot[teal,mark=o, mark size=1.5pt,mark repeat=6, mark phase=6, very thin] table[x index = 1, y index = 0 ]{Data/cdfs/50m_music.tex};
    \addplot[blue,mark=o, mark size=1.5pt,mark repeat=6, mark phase=4, very thin] table[x index = 1, y index = 0 ]{Data/cdfs/50m_esprit.tex};
    \addplot[red,mark=o, mark size=1.5pt,mark repeat=6, mark phase=2, very thin] table[x index = 1, y index = 0 ]{Data/cdfs/50m_2desprit.tex}; 
    \addplot[teal,mark=x, mark size=1.5pt,mark repeat=6, mark phase=6, very thin] table[x index = 1, y index = 0 ]{Data/cdfs/80m_music.tex};
    \addplot[blue,mark=x, mark size=1.5pt,mark repeat=6, mark phase=4, very thin] table[x index = 1, y index = 0 ]{Data/cdfs/80m_esprit.tex};
    \addplot[red,mark=x, mark size=1.5pt,mark repeat=6, mark phase=2, very thin] table[x index = 1, y index = 0 ]{Data/cdfs/80m_2desprit.tex}; 
    \addlegendentry{MUSIC - 15 m}
    \addlegendentry{ESPRIT - 15 m}
    \addlegendentry{2D ESPRIT - 15 m}
    \addlegendentry{MUSIC - 50 m}
    \addlegendentry{ESPRIT - 50 m}
    \addlegendentry{2D ESPRIT - 50 m}
    \addlegendentry{MUSIC - 80 m}
    \addlegendentry{ESPRIT - 80 m}
    \addlegendentry{2D ESPRIT - 80 m}
\end{axis}
\end{tikzpicture}
\caption{CDFs for AoA estimation errors at 15 m, 50 m, and 80 m.}
\label{CDF_TB}
\end{figure}
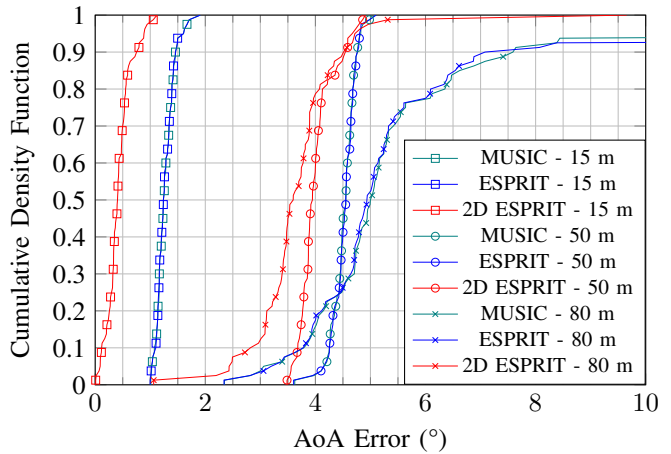
\begin{table}[!tb]
\centering
\caption{Testbed performance for 90\% of CDF\vskip6pt}
\label{Testbed Performance}
\footnotesize
\begin{tabular}{lccc}\toprule
\multirow{2}{*}{\centering Algorithm} & \multicolumn{3}{c}{Distance (m)} \\
& 15 & 50 & 80 \\ 
\midrule
MUSIC      & 1.46\degree{} & 4.77\degree{} & 7.59\degree{} \\
ESPRIT     & 1.46\degree{} & 4.76\degree{} & 7.08\degree{} \\
2D ESPRIT  & 0.75\degree{} & 4.54\degree{} & 4.51\degree{} \\
\bottomrule
\end{tabular}
\label{90cdf}
\end{table}
Figs.~\ref{AOA_TB} and~\ref{CDF_TB} present the AoA estimation results for MUSIC, ESPRIT, and 2D ESPRIT at distances of 15 m, 50 m, and~80~m, along with their CDFs. Table~\ref{90cdf} summarizes the AoA estimation errors in the 90\% CDF for each algorithm. All three methods deliver high accuracy at short distances, but performance deteriorates as the UE–-BS separation increases. This trend is observed consistently in both the AoA time-domain traces and the CDFs. Although MUSIC and ESPRIT perform similarly, 2D ESPRIT consistently shows a clear advantage at all distances, particularly at long ranges, due to its reliance on the full CSI matrix and enhanced spatial resolution. Moreover, in every case biases in the AoA estimation are observed, largely due to our uncalibrated antenna array, the impact of which was shown in Section~\ref{antenna_errors}.

Comparison of the measured results with the simulations in Figs.~\ref{music vs esprit} and~\ref{cal_err} reveals a noticeable discrepancy. The raw model---based on an UMi channel with AWGN and omitting factors like antenna imperfections---does not fully capture the observed performance. To better align simulation outcomes with empirical data, more advanced models are needed that account for a wider range of parameters. Furthermore, although measurements were conducted in an open field, factors such as potential Tx-Rx misalignment at 0\degree{}, terrain height mismatches, and the limited capability of the three-element ULA to resolve more than two sources---which can degrade performance in multipath and interference environments---must be considered.

\section{Conclusion}
\label{conclusions}
This paper presented a roadmap for algorithmic selections in designing a single-anchor 5G UL positioning testbed using USRPs. Candidate algorithms for channel order and AoA estimation were evaluated in simulations before finalizing the design. The impact of antenna calibration errors on AoA estimation was assessed, by modeling gain and phase errors, as well as mutual coupling. In addition, the testbed was evaluated under real conditions for a static UE. Results' analysis shows that while conventional methods such as MUSIC and ESPRIT deliver satisfactory AoA estimates at short distances, their performance deteriorates as the UE-BS distance increases. In contrast, 2D ESPRIT offers significant improvements, particularly at longer distances, due to its utilization of the full CSI matrix and its ability to jointly estimate angle and delay. The discrepancies observed between simulation models and experimental measurements highlight the need to include sophisticated channel and hardware error modeling to improve the predictability of system performance.

\section{Acknowledgments}
The undertaken efforts were conducted within the framework of the Single Node Positioning Testbed (SINGPOS) project funded by the European Space Agency (ESA).

\bibliographystyle{IEEEtran}
\bibliography{IEEEabrv,pimrc}

\end{document}